\documentclass[a4paper,12pt]{article}
\usepackage{citeref}

\input cohmacros.sty

\def\at#1{} 

\advance\textheight2.7cm        
\advance\topmargin-2.0cm
\advance\textwidth2.6cm
\advance\evensidemargin-1.6cm
\advance\oddsidemargin-1.6cm

\begin{document}

\vspace*{-2cm}
\begin{center}
{\LARGE \bf Foundations of quantum physics} \\[4mm]

{\LARGE \bf I. A critique of the tradition} \\

\vspace{1cm}

\centerline{\sl {\large \bf Arnold Neumaier}}

\vspace{0.5cm}

\centerline{\sl Fakult\"at f\"ur Mathematik, Universit\"at Wien}
\centerline{\sl Oskar-Morgenstern-Platz 1, A-1090 Wien, Austria}
\centerline{\sl email: Arnold.Neumaier@univie.ac.at}
\centerline{\sl \url{http://www.mat.univie.ac.at/~neum}}

\end{center}


\hfill April 24, 2019

\vspace{0.5cm}

\bigskip
\bfi{Abstract.}
This paper gives a thorough critique of the foundations of quantum 
physics in its mainstream interpretation (i.e., treating pure states 
and probability as primitives, without reference to hidden variables, 
and without modifications of the quantum laws).

This is achieved by cleanly separating a concise version of the
(universally accepted) formal core of quantum physics from the
(controversial) interpretation issues.
The latter are primarily related to measurement, but also to questions
of existence and of the meaning of basic concepts like 'state' and
'particle'. The requirements for good foundations of quantum physics
are discussed.

Main results:
\\
\pt Born's rule cannot be valid universally, and must be considered as
a scientific law with a restricted domain of validity. 
\\
\pt If the state of every composite quantum system contains all
information that can be known about this system, it cannot be a pure
state in general.

\vfill
For the discussion of questions related to this paper, please use
the discussion forum \\
\url{https://www.physicsoverflow.org}.

\newpage
\vspace*{-2.5cm}
\tableofcontents 


\newpage
\section{Introduction}

This paper gives a thorough critique of the foundations of quantum 
physics in its mainstream interpretation (i.e., treating pure states 
and probability as primitives, without reference to hidden variables, 
and without modifications of the quantum laws).

This is achieved by cleanly separating a concise version of the
(universally accepted) formal core of quantum physics (described in
Section \ref{s.trad}) from the (controversial) interpretation issues.
The latter are primarily related to measurement, but also to questions
of existence and of the meaning of basic concepts like 'state' and
'particle'.

The bridge between the formal core and measurement is Born's rule,
usually assumed to be valid exactly. It is argued in Section
\ref{s.critique} that this assumption cannot be maintained, so that
Born's rule must be considered as a scientific law with a restricted
domain of validity. We also show that if the state of a composite
quantum system contains all information that can be known about a
system, it cannot be a pure state. These are the main new results of
the present paper; they open new possiilities for the foundation of
quantum physics.

Section \ref{s.req} discusses requirements for a foundation of quantum
physics free of the major shortcomings of the traditional
interpretations. The final Section \ref{s.outlook} gives a preview on
the alternative foundation discussed in later parts of this series
\cite{Neu.IIfound,Neu.IIIfound,Neu.IVfound,Neu.Vfound}, that satisfies 
these requirements.

To separate the formal core of quantum physics from the interpretation 
issues we need to avoid some of the traditional quantum mechanical 
jargon. In particular, following the convention of \sca{Allahverdyan} 
et al. \cite{AllBN2}, we add the prefix ''q-'' to all traditional 
quantum notions that suggest by their name a particular interpretation 
and hence might confuse the borderline between theory and 
interpretation. In particular, the Hermitian operators usually called
''observables''\footnote{
This notion appears first in \sca{Dirac}'s 1930 book \cite[p.28]{Dir},
where however, Hermiticity is not required. Later editions make this
restriction.
} 
will be called ''q-observables'' to distinguish them from observables
in the operational sense of numbers obtainable from observation.
Similarly, we use at places the terms q-expectation and q-probability 
for the conventional but formally defined terms expectation and 
probability.

A number of remarks are addressed to experts and then refer to
technical aspects explained in the references given. However, the bulk
of this paper is intended to be nontechnical and understandable for a
wide audience being familiar with some traditional quantum mechanics.
The knowledge of some basic terms from functional analysis is
assumed; these are precisely defined in many mathematics books.

In the bibliography, the number(s) after each reference give the page 
number(s) where it is cited.

\bigskip
{\bf Acknowledgments.}
Earlier versions of this paper benefitted from discussions with
Rahel Kn\"opfel and Mike Mowbray. Thanks also for discussions with
Francois Ziegler about the history of Born's rule.

\section{The traditional foundations of quantum physics}
\label{s.trad}

Quantum physics consists of a \bfi{formal core} that is universally
agreed upon (basically being a piece of mathematics with a few meager
pointers on how to match it with experimental reality) and an
interpretational halo that remains highly disputed even after more than
90 years of modern quantum physics. The latter is the subject of the
\bfi{interpretation of quantum mechanics}, where many interpretations
coexist and compete for the attention of those interested in what
quantum physics really means.

To set the stage, I give here an axiomatic introduction to the
undisputed formal core.

As in any axiomatic setting (necessary for a formal discipline),
there are a number of different but equivalent sets of axioms
or postulates that can be used to define formal quantum physics.
Since they are equivalent, their choice is a matter of convenience.

My choice presented here is the formulation which gives most direct
access to statistical mechanics but is free from allusions to
measurement. The reason for the first is that statistical mechanics is
the main tool for applications of quantum physics to the macroscopic
systems we are familiar with. The reason for the second is that real
measurements consitute a complex process involving macroscopic
detectors, hence should be explained by quantum statistical mechanics
rather than be part of the axiomatic foundations themselves. (This is
in marked contrast to other foundations, and distinguishes the present
system of axioms.)

The relativistic case is outside the scope of the axioms to be
presented, as it must be treated by quantum field theory (see the
discussion in Part II \cite{Neu.IIfound} of this sequence of papers).
Thus the axioms described in the following present nonrelativistic
quantum statistical mechanics in the Schr\"odinger picture.
(As explained later, the traditional starting point is instead the
special case of this setting where all states are assumed to be pure.)

A second reason for my choice is to emphasise the similarity of
quantum mechanics and classical mechanics. Indeed, the only difference
between classical and quantum mechanics in the axiomatic setting
presented below is the following:

\pt
The classical case only works with diagonal operators, where all
operations happen pointwise on the diagonal elements. Thus
multiplication is commutative, and one can identify operators and
functions. In particular, the density operator degenerates into a
probability density.

\pt
The quantum case allows for noncommutative operators, hence both
q-observables and the density are (usually infinite-dimensional)
matrices.

\subsection{Postulates for the formal core of quantum physics}
\label{ss.postulates}

Traditional quantum physics is governed by the following six axioms 
(A1)--(A6).\footnote{
The statements of my axioms contain in parentheses some additional
explanations that, strictly speaking, are not part of the axioms but
make them more easily intelligible; the list of examples given only
has illustrative character and is far from being exhaustive.
} 
Note that all axioms are basis-independent.  $\hbar$ is Planck's
constant, and is often set to 1.

\bfi{(A1)} 
A generic system (e.g., a 'hydrogen molecule') is defined by
specifying a Hilbert space $\Hz$ and a densely defined,
self-adjoint\footnote{\label{f.1}
Self-adjoint operators are Hermitian, $H^*=H$.
Hermitian operators have a real spectrum if and only if they are
self-adjoint.
Hermitian trace class operators are always self-adjoint.
The Hille--Yosida theorem says that $e^{iX}$ exists (and is unitary)
for a Hermitian operator $X$ if and only if $X$ is self-adjoint;
see \sca{Thirring} \cite{Thi3} or \sca{Reed \& Simon} \cite{ReeS.1}.
} 
linear operator H called the \bfi{Hamiltonian} or the
\bfi{internal energy}.

\bfi{(A2)}
A particular system (e.g., 'the ion in the ion trap on this
particular desk') is characterized by its \bfi{state} $\rho(t)$
at every time $t \in\Rz$ (the set of real numbers).
Here $\rho(t)$ is a Hermitian, positive semidefinite, linear trace
class operator on $\Hz$ satisfying at all times the normalization
condition
\[
\Tr\rho(t) = 1.
\]
Here $\Tr$ denotes the trace.

\bfi{(A3)}
A system is called \bfi{closed} in a time interval $[t_1,t_2]$
if it satisfies the evolution equation
\lbeq{e.vN}
\frac{d}{dt}\rho(t) = \frac{i}{\hbar} [\rho(t),H] \for t \in [t_1,t_2],
\eeq
and \bfi{open} otherwise.
If nothing else is apparent from the context, a system is assumed to
be closed.

\bfi{(A4)}
Besides the internal energy $H$, certain other densely defined,
self-adjoint operators or vectors of such operators are
distinguished as \bfi{quantum observables}, short \bfi{q-observables}.
(E.g., the q-observables for a system of $N$ distinguishable particles
conventionally include for each particle several 3-dimensional vectors:
the \bfi{position} $\x^a$, \bfi{momentum} $\p^a$, \bfi{orbital angular 
momentum} $\LL^a$ and the \bfi{spin vector} (or \bfi{Bloch vector})
$\SS^a$ of the particle with label $a$. If $\uu$ is a 3-vector of unit
length then $\uu \cdot \p^a$, $\uu \cdot \LL^a$
and $\uu \cdot \SS^a$ define the momentum, orbital angular momentum,
and spin of particle a in direction $\uu$.)

\bfi{(A5)}
For any particular system, and for every vector $X$ of self-adjoint
q-observables with commuting components, one associates a 
time-dependent monotone linear functional $\<\cdot\>_t$ defining the 
\bfi{q-expectation}
\[
      \<f(X)\>_t:=\Tr\rho(t) f(X)
\]
of bounded continuous functions $f(X)$ at time $t$.
(By \sca{Whittle} \cite{Whi}, this is equivalent to a multivariate
probability measure $d\mu_t(X)$ on a suitable $\sigma$-algebra over the
spectrum $\spec X$ of $X$ defined by
\[
\int d\mu_t(X) f(X) := \Tr\rho(t) f(X) =\<f(X)\>_t.
\]
This $\sigma$-algebra is uniquely determined, and defines
\bfi{q-probabilities}.)

\bfi{(A6)} 
Quantum mechanical predictions consist of predicting properties
(typically q-expecta\-tions or conditional q-probabilities) of the
measures defined in Axiom (A5), given reasonable assumptions about the
states (e.g., ground state, equilibrium state, etc.)

Axiom (A6) specifies that the formal content of quantum physics is
covered exactly by what can be deduced from Axioms (A1)--(A5) without
anything else added -- except for restrictions defining the specific
nature of the states and q-observables, for example specifying
commutation or anticommutation relations between some of the
distinguished q-observables). Thus Axiom (A6) says that Axioms 
(A1)--(A5) are complete.

The description of a particular closed system is therefore given by
the specification of a particular Hilbert space (in (A1)), the
specification of the q-observables (in (A4)), and the
specification of conditions singling out a particular class of
states (in (A6)).
(The description of an open system involves, in addition, the
specification of details of the dynamical law.)

Given this, everything predictable in principle about the system is
determined by the theory, and hence is predicted by the theory.

At the level of the formal core, q-expectations and all other concepts 
introduced are only calculational tools that enables one to predict 
numerical values for theoretical objects with suggestive names. Their 
precise relation to experimental reality is not specified by the formal 
core. These relations are the subject matter of the interpretation of 
quantum mechanics; see Subsection \ref{ss.intCore}.

\subsection{The pure state idealization}\label{ss.pure}

A state $\rho$ is called \bfi{pure} at time $t$ if $\rho(t)$ maps the
Hilbert space $\Hz$ to a 1-dimensional subspace, and \bfi{mixed}
otherwise.

Although much of traditional quantum physics is phrased in terms of
pure states, this is a very special case; in most actual experiments
the systems are open and the states are mixed states. Pure states
are relevant only if they come from the ground state of a
Hamiltonian in which the first excited state has a large energy gap.
Indeed, assume for simplicity that $H$ has a discrete spectrum. In an
orthonormal basis of eigenstates $\phi_k$, functions $f(H)$ of the
Hamiltonian $H$ are defined by
\[
   f(H) = \sum_k f(E_k) \phi_k \phi_k^*
\]
whenever the function $f$ is defined on the spectrum.
The equilibrium density is the canonical ensemble,
\[
\rho(T) = Z(T)^{-1} e^{-H/\kbar T}
= Z(T)^{-1}\sum_k e^{-E_k/T}\phi_k\phi_k^*;
\]
here $\kbar$ is the \bfi{Boltzmann constant}.
(Of course, equating this ensemble with equilibrium in a closed system
is an additional step beyond our system of axioms, which would require
justification.) Since the trace equals 1, we find
\[
   Z(T) = \sum_k e^{-E_k/\kbar T},
\]
the textbook formula for the so-called \bfi{partition function}.
In the limit $T \to 0$, all terms $ e^{-E_k/T}$ become 0 or 1,
with 1 only for the $k$ corresponding to the states with least energy.
Thus, if the ground state $\phi_1$ is unique,
\[
   \lim_{T\to0} \rho(T) = \phi_1 \phi_1^*.
\]
This implies that for low enough temperatures, the equilibrium state
is approximately pure. The larger the gap to the second smallest
energy level, the better is the approximation at a given nonzero
temperature. In particular, the approximation is good if the energy gap
exceeds a small multiple of $E^* := \kbar T$.

States of sufficiently simple systems (i.e., those with a few energy
levels only) can often be prepared in nearly pure states, by realizing
a source governed by a Hamiltonian in which the first excited state has
a much larger energy than the ground state. Dissipation then brings the
system into equilibrium, and as seen above, the resulting equilibrium
state is nearly pure. Those low lying excited states for which a
selection rule suppresses the transition to a lower energy state can be 
made nearly pure in the same way.

\subsection{Schr\"odinger equation and Born's rule}

To see how the more traditional setting in terms of the
Schr\"odinger equation arises, we consider the special case of a closed
system in a pure state $\rho(t)$ at some time $t$. The \bfi{state
vector} of such a system at time $t$ is by definition a unit vector
$\psi(t)$ in the range of the pure state $\rho(t)$. It is determined up
to a phase factor (of absolute value 1), and one easily verifies that
\lbeq{e.pure}
   \rho(t) = \psi(t)\psi(t)^*.
\eeq
Remarkably, under the dynamics for a closed system specified in the
above axioms, this property persists with time if the system
is closed, and the state vector satisfies the 
\bfi{time-dependent Schr\"odinger equation}
\[
  i \hbar \dot\psi(t) = H \psi(t)
\]
Thus the state remains pure at all times. Conversely, for every pure
state, the phases of $\psi(t)$ at all times $t$ can be chosen such that
the Schr\"odinger equation holds; the density operator is independent
of this phase.

Moreover, if $X$ is a vector of q-observables with commuting components
and the spectrum of $X$ is discrete, then the measure from Axiom (A5)
is discrete,
\[
   \int d\mu(X) f(X) = \sum_k p_k f(X_k)
\]
with spectral values $X_k$ and nonnegative numbers $p_k$ summing to 1,
called \bfi{q-probabilities}.\footnote{
This leaves open the precise physical meaning of q-probabilities, and
the question of how to measure them. This is the price to pay for not
entering into interpretational issues.
} 
Associated with the $p_k$ are eigenspaces $\Hz_k$ such that
\[
   X \psi = X_k \psi      \for  \psi \in \Hz_k,
\]
and $\Hz$ is the direct sum of the $\Hz_k$. Therefore, every state
vector $\psi$ can be uniquely decomposed into a sum
\[
   \psi = \sum_k \psi_k,~~~ \psi_k \in \Hz_k.
\]
$\psi_k$ is called the \bfi{projection} of $\psi$ to the eigenspace
$\Hz_k$.
If all eigenvalues of $X$ are discrete and nondegenerate, each $\Hz_k$
is 1-dimensional and spanned by a normalized eigenvector $\phi_k$
Then $X\phi_k=X_k\phi_k$ and the projection is given by
$\psi_k=P_k\psi$ with the orthogonal projector
$P_k:=\phi_k\phi_k^*$, so that
\[
\psi =  \sum_k \phi_k\phi_k^*\psi.
\]
A short calculation using Axiom (A5) now reveals that for a pure state
\gzit{e.pure}, the q-probabilities $p_k$ are given by the
\bfi{formal Born rule}
\lbeq{e.Born0}
   p_k = |\psi_k(t)|^2=|\phi_k^*\psi(t)|^2,
\eeq
where $\psi_k(t)$ is the projection of $\psi(t)$ to the eigenspace
$\Hz_k$.

Identifying these q-probabilities with the probabilities of measurement
results (which is already an interpretative step involving (MI) below)
constitutes the so-called \bfi{Born rule}. Without this identification,
the formal Born rule \gzit{e.Born0} is just a piece of uninterpreted
mathematics with suggestive naming.

Deriving the formal Born rule \gzit{e.Born0} from Axioms (A1)--(A5) 
makes it feel completely natural, while the traditional approach 
starting with Born's rule makes it an irreducible rule full of mystery 
and only justifiable by its miraculous agreement with certain 
experiment.

\subsection{Interpreting the formal core}\label{ss.intCore}

In addition to the formal axioms (A1)--(A6), one needs a rudimentary
interpretation relating the formal part to experiments.
The following \bfi{minimal interpretation} seems to be universally
accepted.

\bfi{(MI)} Upon measuring at times $t_l$ ($l=1,...,n$) a vector $X$ of
q-observables with commuting components, for a large collection of
independent identical (particular) systems closed for times $t<t_l$,
all in the same state
\[
\rho(t_l)=\rho   ~~~(l=1,...,n)
\]
(one calls such systems \bfi{identically prepared}), the measurement
results are statistically consistent with independent realizations
of a random vector $X$ with measure as defined in axiom (A5).

Note that (MI) is no longer a formal statement since it neither defines
what \bfi{measuring} is, nor what \bfi{measurement results} are and what
\bfi{statistically consistent} or \bfi{independent identical system}
means.
Thus (MI) has no mathematical meaning -- it is not an axiom, but already
part of the interpretation of formal quantum physics.

(MI) relates the axioms to a nonphysical entity, the social conventions
of the community of physicists. The terms 'measuring', 'measurement
 results', and 'statistically consistent' already have informal
meaning in the reality as perceived by a physicist. Everything stated
in Axiom (MI) is understandable by every trained physicist.
Thus statement (MI) is not an axiom for formal logical reasoning but
a bridge to informal reasoning in the traditional cultural setting
that defines what a trained physicist understands by reality.

The lack of precision in statement (MI) is on purpose, since it allows
the statement to be agreeable to everyone in its vagueness; different
philosophical schools can easily fill it with their own understanding
of the terms in a way consistent with the remainder.

Interpretational axioms necessarily have this form, since they must
assume some unexplained common cultural background for perceiving
reality. (This is even true in pure mathematics, since the language
stating the axioms must be assumed to be common cultural background.)

(MI) is what {\em every} traditional interpretation I know of assumes
at least implicitly in order to make contact with experiments.
Indeed, all  traditional interpretations I know of assume much more,
but they differ a lot in what they assume beyond (MI).

However, my critique of the universal Born rule in Subsection
\ref{ss.BornLim} also applies to (MI), since (MI) implies the
universal Born rule. Thus (MI) seems to be justified only for certain
measurements.

\bfi{Everything beyond (MI) seems to be controversial}; cf. 
\sca{Schlosshauer} cite[Chapter 8]{Schl,Schl.book}. In particular,
already what constitutes a measurement of $X$ is controversial.
(E.g., reading a pointer, different readers may get marginally
different results. What is the true pointer reading? Does passing a
beam splitter or a polarization filter constitute a measurement?)

On the other hand there is an informal consensus on how to perform
measurements in practice. Good foundations including a good measurement
theory should be able to properly justify this informal consensus by
defining additional formal concepts about what constitutes measurement.
To be satisfying, these must behave within the theory just as their
informal relatives with the same name behave in reality.
The goal of the \bfi{thermal interpretation} described in Part II
\cite{Neu.IIfound} and applied to measurement in Part III 
\cite{Neu.IIIfound} and  Part IV \cite{Neu.IVfound} is to provide such 
foundations.
Its link of the formal core to experiment is different from (MI), based
instead on the link between quantum physics and thermodynamics known 
from statistical mechanics.

\section{A critique of Born's rule}\label{s.critique}

Traditionally, some version of Born's rule
is considered to be an indispensable part of any interpretation of
quantum mechanics, either as a postulate, or as a result derived from
other postulates, not always on the basis of convincing reasoning.
In this section, we have a close look at the possible forms of Born's
rule and discuss the limits of its validity.

\bigskip

All traditional foundations of quantum mechanics heavily depend on the
concept of (hypothetical, idealized) experiments -- far too heavily.
This is one
of the reasons why these foundations are still unsettled, over 90 years
after the discovery of the basic equations for modern quantum mechanics.
No other theory has such controversial foundations.

The main reason is that the starting point of the usual interpretations
is an idealization of the measurement process that is taken too
seriously, namely as the indisputable truth about everything measured.
But in reality, this idealization is only a didactical trick for
the newcomer to make the formal definitions of quantum mechanics a bit
easier to swallow. Except in a few very simple cases, it is too far
removed from experimental practice to tell much about real measurement,
and hence about how quantum physics is used in real applications.

In experimental physics, measurement is a very complex thing -- far
more complex than Born's rule (the usual starting point) suggests.
To measure the distance between two galaxies, the mass of the top
quark, or the Lamb shift -- just to mention three basic examples -
cannot be captured by the idealistic measurement concept used there,
nor by any of the refinements of it discussed in the literature.

In each of the three cases mentioned, one assembles a lot of auxiliary
information and ultimately calculates the measurement result from a
best fit of a model to the data. Clearly the theory must already be in
place in order to do that. We do not even know what a top quark should
be whose mass we are measuring unless we have a theory that tells us
this!

The two most accurately determined observables in the history of
quantum physics, namely the anomalous magnetic moment of the electron
and Lamb shift, are not even q-observables!

To present the stage for the criticism of Born's rule in Subsection 
\ref{ss.BornLim}, and related criticism in Subsections 
\ref{ss.BornVal}-- \ref{ss.pureQFT}, we first need to clarify the 
meaning of the term ''Born's rule''. To distinguish different useful 
meanings we look in Subsections \ref{ss.BornEarly}-- \ref{ss.BornMeas} 
at the early history of Born's rule.

\subsection{Early, measurement-free formulations of Born's rule}
\label{ss.BornEarly}

It is interesting to consider the genesis of Born's rule,\footnote{
apparently named such first in 1934 by \sca{Bauer} \cite[p.302]{Bau34}
({\it ''la r\`egle de Born''}). Before that, during the gestation
period of finding the right level of generalization and interpretation,
the pioneers talked more vaguely about Born's interpretation of quantum
mechanics (or of the wave function). For example, \sca{Jordan}
\cite[p.811]{Jor1927} writes about {\it ''Born's Deutung der L\"osung
[der] Schr\"odingergleichung''}.
} 
based on the early papers of the pioneers of quantum mechanics.
This and the next subsection benefitted considerably from discussions
with Francois Ziegler, though his view of the history is somewhat 
different (cf. \sca{Ziegler} \cite{Zie}).

The two 1926 papers by \sca{Born} \cite{Bor1926a,Bor1926b} (the first
being a summary of the second) introduced the probabilistic
interpretation that earned Born the 1954 Nobel prize.  Born's 1926
formulation\footnote{
German original,  \sca{Born} \cite[p.865f]{Bor1926a}:
{\it ''bestimmt die Wahrscheinlichkeit daf\"ur, da{\ss} das aus der
$z$-Richtung kommende Elektron in die durch $\alpha,\beta,\gamma$
bestimmte Richtung (und mit einer Phasen\"anderung $\delta$) geworfen
wird''}
} 
{\it ''gives the probability for the electron, arriving from the
$z$-direction, to be thrown out into the direction designated by the
angles $\alpha, \beta, \gamma$, with the phase change $\delta$''}
does not depend on anything being measured, let alone to be assigned a
precise numerical measurement value! Instead it sounds like talk about
objective properties of electrons (''being thrown out'') independent of
measurement. Thus Born originally did not relate his interpretation to
measurement but to objective properties of scattering processes, no
matter whether these were observed.

Rephrased in modern terminology (Born didn't have the concept of an
S-matrix), Born's statement above is made precise (and generalized)
by the following rule:

\bfi{Born's rule (scattering form)}:
In a scattering experiment described by the S-matrix $S$,
\[
\Pr(\psi_\out|\psi_\iin):=|\psi_\out^*S\psi_\iin|^2
\]
is the conditional probability density that scattering of particles
prepared in the in-state $\psi_\iin$ results in particles
in the out-state $\psi_\out$. Here the in- and out-states
are asymptotic eigenstates of total momentum, labelled by a maximal
collection of independent quantum numbers (including particle momenta
and spins).

The scattering form of Born's rule is impeccable and remains until
today the basis of the interpretation of S-matrix elements computed
from quantum mechanics or quantum field theory.

\bigskip

The 1927 paper by \sca{Born} \cite{Bor1927} extends this rule on p.173
to probabilities for quantum jumps (''Quantensprung'', p.172)
between energy eigenstates, given by the absolute squares of inner
products of the corresponding eigenstates, still using objective rather
than measurement-based language:\footnote{
{\it ''Das Quadrat $|b_{nm}|^2$ ist gem\"a{\ss} unserer Grundhypothese
die Wahrscheinlichkeit daf\"ur, da{\ss} das System sich nach Ablauf der
St\"orung im Zustand $m$ befindet''}
} 
For a system initially in state $n$ given by Born's (9),
{\it ''the square $|b_{nm}|^2$ is according to our basic hypothesis
the probability for the system to be in state $m$ after completion of
the interaction''}.
Here state $n$ is the $n$th stationary state (eigenstate with a
time-dependent harmonic phase) of the Hamiltonian.

Born derives this rule from two assumptions. The first assumption, made
on p.170 and repeated on p.171 after (5), is that an atomic system is
always in a definite stationary state:\footnote{
{\it ''Wir werden also an dem Bohrschen Bilde festhalten, da{\ss} ein
atomares System stets nur in einem stationaren Zustand ist. [...]
im allgemeinen aber werden wir in einem Augenblick nur wissen,
da{\ss} auf Grund der Vorgeschichte und der bestehenden physikalischen
Bedingungen eine gewisse Wahrscheinlichkeit daf\"ur besteht, da{\ss} das
Atom im $n$-ten Zustand ist.''}
} 
''{\it Thus we shall preserve the picture of Bohr that an atomic system
is always in a unique stationary state. [...]
but in general we shall know in any moment only that, based on
the prior history and the physical conditions present, there is a
certain probability that the atom is in the $n$th state.''}

Thus for the early Born, the beables of a quantum mechanical system are
the quantum numbers of the (proper or improper) stationary states of the
system. This assumption works indeed for equilibrium quantum
statistical mechanics -- where expectations are defined in terms of the
partition function and a probability distribution over the stationary
states. It also works for nondegenerate quantum scattering theory
-- where only asymptotic states figure. However, it has problems in the
presence of degeneracy, where only the eigenspaces, but not the
stationary states themselves, have well-defined quantum numbers.
Indeed, \cite{Bor1927} assumes -- on p.159, remark after (2) and
Footnote 2 -- that the Hamiltonian has a nondegenerate, discrete
spectrum.

Born's second assumption is his basic hypothesis on p.171 for
probabilities for being (objectively) in a stationary state:\footnote{
{\it ''[...] eine gewisse Wahrscheinlichkeit daf\"ur besteht, da{\ss}
das Atom im $n$-ten Zustand ist. Wir behaupten nun, da{\ss} als Ma{\ss}
dieser Zustandswahrscheinlichkeit die Gr\"o{\ss}e
$|c_n|^2=|\int \psi(x,t)\psi_n^*(x)dx|^2$ zu w\"ahlen ist.''}
} 
{\it ''there is a certain probability that the atom is in the $n$th
state. We now claim that as measure for this probability of state, one
must choose the quantity $|c_n|^2=|\int \psi(x,t)\psi_n^*(x)dx|^2$.''}

The 1927 paper by \sca{Jordan} \cite[p.811]{Jor1927} (citing Pauli)
extends Born's second assumption further to an objective, measurement
independent probability interpretation of inner products (probability
amplitudes) of eigenstates of two arbitrary operators, seemingly without
being aware of the conceptual problem this objective view poses when
applied to noncommuting operators.

The 1927 paper by \sca{Pauli} \cite[p.83,Footnote 1]{Pau1927} contains
the first formal statement of a probability interpretation for
position:\footnote{
 {\it ''Wir wollen diese [...] Funktion im Sinne der von Born in seiner
Sto{\ss}mechanik [here he cites \cite{Bor1926a,Bor1926b}] vertretenen
Auffassung des ''Gespensterfeldes'' folgenderma{\ss}en deuten: Es ist
$|\psi(q_1\ldots q_f)|^2dq_1\ldots dq_f$ die Wahrscheinlichkeit daf\"ur,
da{\ss} im betreffenden Quantenzustand des Systems diese Koordinaten
sich zugleich im betreffenden Volumenelement $dq_1\ldots dq_f$ des
Lageraums befinden.''}
} 
{\it ''We shall interpret this function in the spirit of Born's view of
the ''Gespensterfeld'' in \cite{Bor1926a,Bor1926b} as follows:
$|\psi(q_1\ldots q_f)|^2dq_1\ldots dq_f$ is the probability that,
in the named quantum state of the system, these coordinates lie
simultaneously in the named volume element  $dq_1\ldots dq_f$ of
position space.''}
Apart from its objective formulation (no reference to measurement),
this is a special case of the universal formulation of Born's rule
given below:

The 1927 paper by \sca{von Neumann} \cite[p.45]{vNeu1927a} generalizes
this statement to arbitrary selfadjoint operators, again stated as an
objective (i.e., measurement independent) interpretation. For discrete
energy spectra and their energy levels, we still read p.48:
{\it ''unquantized states are impossible'' (''nicht gequantelte
Zust\"ande sind unm\"oglich'')}.

Note that like Born, Jordan and von Neumann both talk about objective
properties of the system independent of measurement. But unlike Born
who ties these properties to the stationary state representation in
which momentum and energy act diagonally, Pauli ties it to the
position representation, where position acts diagonally, and von Neumann
allows it for arbitrary systems of commuting selfadjoint operators.

From either Born's or Jordan's statement one can easily obtain the
following, basis-indepen\-dent form of Born's rule, either for functions
$A$ of stationary state labels, or for functions $A$ of position:

\bfi{Born's rule (objective expectation form)}:
The value of a q-observable corresponding to a self-adjoint Hermitian 
operator $A$ of a system in the pure state $\psi$ (or the mixed state 
$\rho$) equals on average the q-expectation value $\<A\>:=\psi^*A\psi$ 
(resp. $\<A\>:=\Tr\rho A$).

The first published statement of this kind seems to be in the 1927 paper
by \sca{Landau} \cite[(4a),(5)]{Lan1927}. The interpretational part
is in Footnote 2 there, which states that (the formula corresponding in
modern notation to $\<A\>:=\Tr\rho A$) denotes the probability mean
(''Wahrscheinlichkeitsmittelwert''). Again there is no reference to
measurement.

\subsection{Formulations of Born's rule in terms of measurement}
\label{ss.BornMeas}

As pointed out by \sca{Weyl} \cite[p.2]{Weyl1927}, the derivation
from Born's and Jordan's statement does not extend to general operators
$A$, due to noncommutativity and the resulting complementarity.
Another consequence of this noncommutativity is that Born's stationary
state probability interpretation and Pauli's position probability
density interpretation cannot both claim objective status.
Therefore, later interpretations relate the notion of value of an
observable (being in the $n$-state, or having position $r$) more 
directly to measurement.

The 1927 paper by \sca{von Neumann} \cite{vNeu1927b},
after having noted (on p. 248) the problems resulting from noncommuting
quantities that cannot be observed simultaneously, derives axiomatically
on p.255 he derives for a theoretical expectation value with natural
properties the necessity of the formula $\<A\>:=\Tr\rho A$ with
Hermitian $\rho$ (his $U$) of trace 1. This is abstract mathematical
reasoning independent of any relation to measurement, and hence belongs
to the formal (uninterpreted) core of quantum physics. However, the
motivation for his axioms, and hence their interpretation, is taken
from a consideration on p.247 of the measurement of values in an
ensemble of systems, taking the expectation to be the ensemble mean
of the measured values. Specialized to a uniform (''einheitlich'')
ensemble of systems in the same completely known (pure) state $\psi$
of norm one he then finds on p.258 that $\rho=\psi\psi^*$, giving
$\<A\>:=\psi^*A\psi$.

In the present terminology, we may phrase von Neumann's interpretation
of q-expectation values as follows:

\bfi{Born's rule (measured expectation form)}:
If a q-observable corresponding to a self-adjoint Hermitian operator 
$A$ is measured on a system in the pure state $\psi$ (or the mixed 
state $\rho$), the results equal on average the q-expectation value 
$\<A\>:=\psi^*A\psi$ (resp. $\<A\>:=\Tr\rho A$).

Note that the q-expectation value has a formal meaning independent of
the interpretation; the measured expectation form of Born's rule just
asserts that measurements result in a random variable whose expectation
agrees with the formal q-expectation. To justify the ''equal'', the
average in question cannot be a sample average (where only an
approximate equal results, with an accuraccy depending on size and
independence of the sample) but must be considered as the theoretical
expectation value of the random variable.

On a purist note, we can only take finitely many measurements on a
system. But the expectation value of a random variable is insensitive
to the result of a finite number of realizations. Thus, in the most
stringent sense, the expectation form of Born's rule says nothing at
all about measurement.
However, the content of the expectation form is roughly the content of
the more carefully formulated statement (MI) discussed in Subsection
\ref{ss.intCore}, specialized to a pure state. (MI) does not have the
defect just mentioned.

More conventionally, Born's rule is phrased in terms of measurement
results and their probabilities rather than expectations.
As part of Born's rule, it is usually stated
(see, e.g., \cite{Wik.Born}) that the results of the measurement of a
q-observable exactly equals one of the eigenvalues.

A precise basic form of Born's rule (often augmented by a more
controversial collapse statement about the state after a
measurement\footnote{
For example, in his famous 1930 book, \sca{Dirac} \cite[p.49]{Dir}
states:
{\it ''The state of the system after the observation must be an
eigenstate of [the observable] $\alpha$, since the result of a
measurement of $\alpha$ for this state must be a certainty.''}
In the third edition \cite[p.36]{Dir3}, he writes:
{\it ''Thus after the first measurement has been made, the system is in
an eigenstate of the dynamical variable $\xi$, the eigenvalue it
belongs to being equal to the result of the first measurement.
This conclusion must still hold if the second measurement is not
actually made. In this way we see that a measurement always causes the
system to jump into an eigenstate of the dynamical variable that is
being measured, the eigenvalue this eigenstate belongs to being equal
to the result of the measurement.''} \\
A 2007 source is \sca{Schlosshauer} \cite{Schl.book}, who takes the
collapse (''jump into an eigenstate'') to be part of what he calls the
''standard interpretation'' of quantum mechanics, but does not count it
as part of Born's rule (p.35). On the other hand,
\sca{Landau \& Lifschitz} \cite[Section 7]{LL.3} explicitly remark that
the state after the measurement is in general not an eigenstate.
} 
not discussed here) is the following, taken almost verbatim from
Wikipedia \cite{Wik.Born}.

\bfi{Born's rule (discrete form)}:
If a q-observable corresponding to a self-adjoint Hermitian operator 
$A$ with discrete spectrum is measured in a system described by a pure 
state with normalized wave function $\psi$ then
\\
(i) the measured result will be one of the eigenvalues $\lambda$ of
$A$, and
\\
(ii) the probability of measuring a given eigenvalue $\lambda_i$ equals
$\psi^*P_i\psi$, where $P_i$ is the projection onto the eigenspace
of $A$ corresponding to $\lambda_i$.

A related statement is claimed to hold for arbitrary spectra with a
continuous part, generalizing both the discrete form and the original
form.

\bfi{Born's rule (universal form)}:\footnote{\label{f.Weinberg}
The quantum field theory book by \sca{Weinberg} \cite{Wei1} pays on p.50
(2.1.7) lip service to the universal form of Born's rule. But the only
place where Born's rule is used is on p.135 (3.4.7), where instead the
scattering form is employed to get the transition rates for scattering
processes. Thus quantum field theory only relies on the scattering form
of Born's rule.
} 
If a q-observable corresponding to a self-adjoint Hermitian operator
$A$ is measured in a system described by a pure state with normalized
wave function $\psi$ then
\\
(i) the measured result will be one of the eigenvalues $\lambda$ of
$A$, and
\\
(ii) for any open interval $\Lambda$ of real numbers, the probability
of measuring $\lambda\in\Lambda$ equals $\psi^*P(\Lambda)\psi$, where
$P(\Lambda)$ is the projection onto the invariant subspace
of $A$ corresponding by the spectral theorem to the spectrum in
$\Lambda$.

If the measurement result $\lambda_i$ is an isolated eigenvalue of $A$,
the universal form reduces to the discrete form, since one can take
$\Lambda$ to be an open interval intersecting the spectrum in
$\lambda_i$ only, and in this case, $P(\Lambda)=P_i$.

Using the spectral theorem, it is not difficult to show that the
universal form of Born's rule implies the measured expectation form.
Conversely, the measured expectation form of Born's rule almost implies
the universal form.
It fully implies the second part (ii), from which it follows that the
 first part (i) holds with probability 1 (but not with certainty, as
the universal form claims).\footnote{\label{f.hair}
This is not just hair splitting. The difference between probability 1
and certainty can be seen by noting that the probability that a random
number drawn uniformly from $[0,1]$ is irrational with probability 1,
while measurements usually produce rational numbers.
\\
In general, the expectation form certainly allows finitely many
exceptions to (i), and hence says, strictly speaking, nothing at all 
about the possible measurement results. (This is well illustrated in 
the story ''The Metaphysian's Nightmare'' by the logician Bertrand 
\sca{Russell} \cite{Rus.Nightmare}: ''There is a special department of 
Hell for students of probability. In this department there are many 
typewriters and many monkeys. Every time that a monkey walks on a 
typewriter, it types by chance one of Shakespeare's sonnets.'') 
However, the application to spin measurements and to quantum 
information theory requires that (i) holds at least for binary spectra. 
Hence the expectation form is slightly deficient.
} 

Unfortunately, the application of Born's rule to measurement problems
in general is highly questionable. Because of the equivalence just
mentioned, it is enough to discuss the universal form of Born's rule.

\subsection{Limitations of Born's rule}\label{ss.BornLim}

Though usually stated as universally valid, Born's rule has severe
limitations. In the universal form, it neither applies to photodetection
nor to the measurement of the total energy, just to mention the most
conspicuous misfits. Moreover, equating the results of measurements
with exact eigenvalues is very questionable when the latter are
irrational or (as in the case of angular momentum) multiples of a not
exactly known constant of nature. In addition, real measurements rarely
produce exact numbers (as Born's rule would require it) but (cf.
\cite{SI2}) numbers that are themselves subject to uncertainty.
Because of these limitations and the inherent ambiguities in specifying
what constitutes a measurement\footnote{
This is a highly nontrivial problem in quantum
statistical mechanics; cf. \sca{Allahverdyan} et al. \cite{AllBN1}.
} 
and what qualifies as a measurement result, subsequent derivations can
never claim universal validity either.

Problems with Born's rule include:

\begin{enumerate}
\item
At energies below the dissociation threshold (i.e., where the spectrum
of the Hamiltonian $H$, the associated q-observable, is discrete),
energy measurements of a system almost never yield an exact eigenvalue
of $H$. For example, nobody knows the exact value of the Lamb shift, a 
difference of eigenvalues of the Hamiltonian of the hydrogen atom; 
the (reasonably) precise measurement was even worth a Nobel prize
(1955 for Willis Lamb). Indeed, the energy levels of most realistic 
quantum systems are only inaccurately known.

\item
In particular, Born's rule does not apply to the total energy of a 
composite system, one of the key q-observables\footnote{
{\it ''we shall assume the energy of any dynamical system to be
always an observable''} (\sca{Dirac} \cite[p.38]{Dir})
} 
in quantum physics, since the spectrum is usually very narrowly spaced
and precise energy levels are known only for the simplest systems in
the simplest approximations. Therefore Born's rule cannot be used to 
justify the canonical ensemble formalism of statistical mechanics; 
it can at best motivate it.

\item
The same holds for the measurement of masses of relativistic particles
with 4-momen\-tum $p$, which never yield exact eigenvalues of the mass
operator $M:=\sqrt{p^2}$. Indeed, the masses of most particles are
only inaccurately known.

\item
When a particle has been prepared in an ion trap (and hence is there
with certainty), Born's rule implies a tiny but positive probability 
that at an arbitrarily short time afterwards it is detected a light year
away.\footnote{\label{f.nonlocalProb}
Indeed, for a single massive particle, Born's rule states that
$|\psi(x,t)|^2$ is the probability density for locating at a given
time $t$ the particle at a particular position $x$ anywhere in the
universe, and the Fourier transform $|\wt\psi(p,t)|^2$ is the
probability density for locating at a given time $t$ the particle with
a particular momentum $p$. In the present case, the position density
has bounded support, so by a basic theorem of harmonic analysis, the
momentum density must have unbounded support. This implies the claim.
} 
In a similar spirit, \sca{Heisenberg} \cite[p.25]{Heisenberg1930}
wrote in 1930:\footnote{
German original: {\em ''Das Resultat ist aber merkw\"urdiger, als es 
im ersten Augenblick den Anschein hat. Bekanntlich nimmt $\psi^*\psi$ 
exponentiell mit wchsendem Abstand vom Atomkern ab. Also besteht immer 
noch eine endliche Wahrscheinlichkeit daf\"ur, das Elektron in sehr 
weitem Abstand vom Atomkern zu finden.''}
} 
{\em ''This result is stranger than it seems at first glance. As is 
well known, $\psi^*\psi$ diminishes exponentially with increasing 
distance from the nucleus; there is thus always a small but finite 
probability of finding the electron at a great distance from the center 
of the atom.''}
Thus $|\psi(x)|^2$ cannot be the exact probability density for being
detected at $x$.

\item
This argument against the exact probability density interpretation of
$|\psi(x)|^2$ works even for relativistic particles in the multiparticle
framework of \sca{Keister \& Polyzou} \cite{KeiP}.

\item
A no-go theorem for exact measurement by \sca{Wigner} \cite{Wig.nogo}
rules out projective measurements of a particle being in a given region,
since the corresponding projector does not commute with all additive
conserved quantities. See also \sca{Ozawa} \cite{Oza} and
\sca{Araki \& Yanase} \cite{AraY}.

\item
Many measurements in quantum optics are POVM measurements
\cite{wik.POVM}, i.e., described by a positive operator-valued measure.
These follow a different law of which Born's rule is
just a very special case where the POVM is actually projection-valued.
(In general, the POVM law can be derived from Born's rule
applied to a fictitious von Neumann experiment in an extended
Hilbert space. This shows its consistency with Born's rule but still
disproves the latter for the actual physical states in the smaller
physical Hilbert space.)

\item
Born's rule does not cover the multitude of situations where typically
only single measurements of a q-observable are made. In particular,
Born's rule does not apply to typical macroscopic measurements, whose
essentially deterministic predictions are derived from statistical
mechanics.

\item
The measurement of quantum fields is not covered by Born's rule.
These are q-observables depending on a space-time argument, and one can
prepare or measure events at any particular space-time position at most
once. Thus it is impossible to repeat measurements, and the standard
statistical interpretation in terms of sufficiently many identically
prepared systems is impossible.

\item
Many things physicists measure have no simple interpretation in terms
of a Born measurement. Examples include spectral lines and widths,
particle masses and life times, chemical reaction rates, or scattering
cross sections. Often lots of approximate computations are involved.
\end{enumerate}

To uphold Born's rule in view of points 1.-3., one would have to 
consider the concept of measurement that of a fictitious, infinitely 
precise measurement.\footnote{
In particular, this would exclude a subjective interpretation of 
measurement results and associated probabilities in terms of 
the experimenter's knowledge.
} 

Points 4 and 5 above imply that the wave function and hence the
density operator encode in their basic operational interpretation
highly nonlocal information.
Thus nonlocality is explicitly built into the very foundations of
quantum mechanics {\it as conventionally presented}.
Processing nonlocal information, it is no surprise that standard
quantum mechanics defined by the Schr\"odinger equation violates the
conclusions of Bell type theorems (cf. the discussion and references in
\sca{Neumaier} \cite{Neu.ens}).
It already violates their assumptions!

Points 4 and 5 also show that at finite times (i.e., outside its use
to interpret asymptotic S-matrix elements), Born's rule cannot be
strictly true in relativistic quantum field theory, and hence not in
nature.

\subsection{The domain of validity of Born's rule}\label{ss.BornVal}

As we have seen, Born's rule has, like any other statement in physics,
its domain of validity but leads to problems when applied outside this
domain.\footnote{
Progressing from the Born rule to so-called positive operator valued
measures (POVMs) is already a big improvement, commonly used in quantum
optics and quantum information theory. These are adequate for
measurements in the form of clicks, flashes or events (particle tracks)
in scattering experiments, and perhaps only then.
\\
But these still do not cover measurements of energy, or of the Lamb
shift, or of particle form factors.
} 
From an analysis of many different q-observables and
measurement protocols, it seems that the discrete form of Born's rule
needs four conditions for its validity.
It is valid precisely for measuring q-observables\\
1.
with only discrete spectrum,\\
2.
measured over and over again in identical states (to make sense of the
probabilities), where\\
3.
the difference of adjacent eigenvalues is significantly larger than
the measurement resolution, and where\\
4.
the measured value is adjusted to exactly match the spectrum, which
must be known exactly prior to the measurement.

The universal version has similar limitations also when restricted to
purely continuous spectra; in this case it seems to be valid
only in Born's original scattering form.

\subsection{What is a state?}\label{ss.state}

In physics, the state of a physical system (whether classical or
quantum) gives a complete description of the system at a given time.
The following is a concise formulation of this:

{\bf (S1)}
The state of a system (at a given time) encodes everything that can be
said (or ''can be known'') about the system at this time, including
the possible predictions at later times, and nothing else.

Thus every property of the system can (in principle) be computed from
its state.

For a complex system, knowledge about the whole system is usually
obtained by collecting knowledge about its various parts. This makes
sense only if we require in addition,

{\bf (S2)}
Every property of a subsystem is also a property of the whole system.

Indeed, not knowing something about the subsystem means not knowing
everything about the system as a whole, and hence not knowing the
precise state of the system.

Thus common sense dictates that a sound, observer-independent
interpretation of quantum physics should satisfy (S1) and (S2).

Now (S2) says that the state of the full system determines all
properties of any of its subsystems. hence it determines by
(S1) the state of each subsystem to the last detail. Thus we conclude:

{\bf (S3)} The state of a system determines the state of all its
subsystems.

A macroscopic body should therefore have a valid
microscopic quantum description -- a quantum state -- that determines
all observable properties on every level. In particular, an approximate
hydromechanical classical description for the most important observable
properties, namely the q-expectation values of the fields, must be
obtainable from this exact quantum state. (The process to achieve this
is usually called coarse graining.)

Property (S3) must hold for logical reasons even though in practice we
may never know the precise state of the system and/or the subsystems.
Indeed, we usually know only very little information about any system,
unless the latter is so tiny that it can be fully described by very few
parameters.

\bigskip

Unfortunately, none of the mainstream versions of the interpretation
of quantum mechanics (i.e., those not invoking hidden variables) are
anywhere presented in a form that would satisfy our conclusion (S3). 
Since the deficiency always has the same root -- the treatment of
the density operator as representing a state of incomplete knowledge, 
a statistical mixture of pure states -- it is enough to discuss one
specific interpretation. We shall look at the interpretation given in 
the very influential treatise of theoretical physics by 
\sca{Landau \& Lifshitz} \cite{LL.3,LL.5}. They start their discussion 
of quantum mechanics with a particular version of Born's rule:

\cite[p.6]{LL.3} {\it
''The basis of the mathematical formalism of quantum mechanics lies in
the proposition that the state of a system can be described by a
definite (in general complex) function $\Psi(q)$ of the coordinates.
The square of the modulus of this function determines the probability
distribution of the values of the coordinates: $|\Psi^2|dq$ is the
probability that a measurement performed on the system will find the
values of the coordinates to be in the element $dq$ of configuration
space. The function $\Psi$ is called the wave function of the system.
[...]  If the wave function is known at some initial instant, then,
from the very meaning of the concept of complete description of a state,
it is in principle determined at every succeeding instant.''}

Thus in terms of our formal core, the complete description of the
system is declared by Landau \& Lifshitz to be a pure state, and the
properties of the system are declared to be the probabilities of
potential measurement results. They then consider parts (subsystems)
and observe:

\cite[p.7]{LL.3} {\it
''Let us consider a system composed of two parts, and suppose that the
state of this system is given in such a way that each of its parts is
completely described.$\dagger$''}\\
Footnote:  {\it
''$\dagger$ This, of course, means that the state of the whole system
is completely described also. However, we emphasize that the converse
statement is by no means true: a complete description of the state of
the whole system does not in general completely determine the states
of its individual parts''}

Thus they explicitly deny (S3). In fact, except in the special case
 discussed in the context of the above quote -- where the state
factors into a tensor product of states of the subsystems -- {\em they
do not indicate at all how the state of a system and that of its parts
are related.}
More mysteriously, nowhere in the literature seems to be a
discussion that would tell us anything on the formal level about the
relationship between the pure state of a system and the pure state of
a subsystem. There seems to be no such relation, except in the
idealized, separable case mentioned above, usually assumed to be valid
only before an interaction happens.

But this would mean that the quantum state of a physics lab has nothing
to do with the quantum states of the equipment in it, and of the
particles probed there!
This is {\em very} strange for a science such as physics that studies
large systems primarily by decomposing them into its simple constituents
and infers properties of the former from collective properties of the
latter.

This truly unacceptable situation shows that there is something
deeply wrong with the traditional interpretations.\footnote{
In addition, there are the traditional difficulties of interpretations
of quantum mechanics, well summarized in Section 3.7 of the quantum
mechanics textbook by \sca{Weinberg} \cite{Wei.QM}.
} 
It is no surprise that this leads to counterintuitive paradoxes in
situations -- such as experiments with entangled photons -- where a
larger (e.g., 2-photon) system is prepared but its constituents (here
2 single photons) are observed.

\bigskip

Of course, we can never know the exact quantum state of a physics lab
or a piece of equipment. Because of that it has become respectable to
interpret quantum mechanics not in terms of what is but in terms of
what is known to the person modeling a physical system. The system state
then becomes a complete description no longer of the physical system
but of the knowledge available.\footnote{
Whose knowledge this is is usually not addressed; presumably it is the
knowledge of the person creating the mathematical model of the quantum
system. Taken at face value, this would make the system state a function
of the mental state of the modeler's mind -- another truly
unacceptable situation.
} 
Uncertainty about the pure state is then modeled as a probability
distribution for being in a pure state. Averaging with corresponding
weight leads to more general mixed states described by density
operators. In this context, \sca{Landau \& Lifshitz} write:

\cite[p.16]{LL.5} {\it
''The quantum-mechanical description based on an incomplete set of data
concerning the system is effected by means of what is called a density
matrix [...] The incompleteness of the
description lies in the fact that the results of various kinds of
measurement which can be predicted with a certain probability from a
knowledge of the density matrix might be predictable with greater or
even complete certainty from a complete set of data for the system,
from which its wave function could be derived.''}

Based on this, they derive the interpretation of the q-expectation
$\<A\>:=\Tr\rho A$ as the expectation value of $A$ in a mixed state
$\rho$:

\cite[p.17]{LL.5} {\it
''The change from the complete to the incomplete quantum-mechanical
description of the subsystem may be regarded as a kind of averaging
over its various $\psi$ states. [...] the mean value $\ol f$ becomes
the trace (sum of diagonal elements) of this operator''}

However, on the next page they call their description an
illustration only, denying it any trace of reality:

\cite[p.18]{LL.5} {\it
''It must be emphasised that the averaging over various $\psi$ states,
which we have used in order to illustrate the transition from a
complete to an incomplete quantum-mechanical description, has only a
very formal significance. In particular, it would be quite incorrect to
suppose that the description by means of the density matrix signifies
that the subsystem can be in various $\psi$ states with various
probabilities and that the averaging is over these probabilities.
Such a treatment would be in conflict with the basic principles of
quantum mechanics.\\
The states of a quantum-mechanical system that are described by wave
functions are sometimes called pure states, as distinct from mixed
states, which are described by a density matrix. Care should, however,
be taken not to misunderstand the latter term in the way indicated
above.\\
The averaging by means of the statistical matrix according to (5.4) has
a twofold nature. It comprises both the averaging due to the
probabilistic nature of the quantum description (even when as complete
as possible) and the statistical averaging necessitated by the
incompleteness of our information concerning the object considered.
For a pure state only the first averaging remains, but in statistical
cases both types of averaging are always present. It must be borne in
mind, however, that these constituents cannot be separated; the whole
averaging procedure is carried out as a single operation,and cannot be
represented as the result of successive averagings, one purely
quantum-mechanical and the other purely statistical.''}

Thus Landau and Lifschitz reject the subjective, knowledge-based view, 
as one cannot divide the information contained in the density operator 
into an objective, pure part corresponding to the objective properties 
of the system and a statistical part accounting for lack of knowledge.

But this also means that their derivation of the interpretation of the
q-expectation $\<A\>:=\Tr\rho A$ as expectation value is invalid, being
based on an invalid illustration only.
Note that this formula is heavily used in quantum statistical mechanics
and quantum field theory. It is often applied there in contexts where
no measurement at all is involved and when it is not even clear how one
should measure the operators in question.\footnote{
We mentioned already in Subsection \ref{ss.BornLim} the problems
with interpreting measurements of the energy, corresponding to the
operator $H$ figuring in all of quantum statistical mechanics.
} 
Indeed, most of quantum statistical mechanics is not concerned with
measurement at all.
In all these cases the connection to measurement and hence to Born's
rule is absent, and even the hand-waving ''illustrative'' derivation
given is spurious, as the items going into the derivation are never
actually measured.

On the other hand, the use of the density operator is central to quantum
statistical mechanics. The fact that the latter predicts qualitatively
and quantitatively the thermodynamics of macroscopic systems shows
that {\bf the density operator contains objective, knowledge-independent
information, and is the true carrier of the state information in
quantum physics.} This is one of the reasons why the description of
the formal core of quantum physics presented in Subsection
\ref{ss.postulates} featured the density operator as basic. Pure states
then appear as idealizing approximation under the conditions discussed
in Subsection \ref{ss.pure}.

It is very remarkable that thermodynamics provides an alternative
interpretation of the q-expectation $\<A\>:=\Tr\rho A$ -- not as as
expectation value, but as the macroscopic, approximately measured value
of $A$. This is the germ of the thermal interpretation of quantum
physics to be discussed in Part II \cite{Neu.IIfound}.

\subsection{Pure states in quantum field theory}\label{ss.pureQFT}

That pure states cannot have a fundamental meaning can also be seen
from the perspective of quantum field theory. It is a very little known
fact that, in any interacting relativistic quantum field theory,
the notion of a pure state loses its meaning. Results from algebraic
quantum field theory (cf. \sca{Yngvason} \cite[p.12]{YngIII}) imply that
all local algebras induced by a relativistic quantum field theory on a
causal diamond (an intersection of a future cone and a past cone with
nonempty interior) are factors of type $III_1$ in von Neumann's
classification of factors as refined by \sca{Connes} \cite{Con}.
Picking such a causal diamond containing our present planetary system
implies that we may
assume the algebra of observables currently accessible to mankind to be
such a factor of type $III_1$. Remarkably, such a $C^*$-algebra $\Az$
has no pure states \cite[p.14]{YngIII}.\footnote{
An explicit example of a factor of type $III_1$ involving an infinite
array of spin-1/2 particles is given in equations (27) and (29) of
\sca{Yngvason} \cite{YngIII}.
} 
Therefore, in these representations, one cannot rigorously argue about
states by considering partial traces in nonexistent pure states!
This shows that pure states must be the result of
a major approximating simplification, and not something fundamental.

Note that $\Az$ has infinitely many unitarily inequivalent irreducible
representations on Hilbert spaces (corresponding to the different
superselection sectors of the theory). But in each such representation,
the algebra $\Az$ of bounded q-observables is vanishingly small compared
to the algebra of all bounded operators. A vector state
in the Hilbert space $\Hz$ of an irreducible representation of a local
algebra $\Az$ of type $III_1$ (which is a pure state in $\Hz$) can
therefore still be mixed as a state of $\Az$.\footnote{
For example, for an arbitrary mixed state of a Hilbert space $\Hz_0$,
the GNS construction produces another Hilbert space in which this state
is pure. Note that in quantum physics, the GNS construction is of
limited value only, as this Hilbert space depends on the state one
started with, while standard quantum mechanics works with pure states
contained in a \bfi{fixed} Hilbert space. One therefore needs a
distinguished state to define a Hilbert space. Now the only 
distinguished state in quantum field theory is the vacuum state. 
But for gauge theories such as quantum electrodynamics, the Hilbert 
space corresponding to this vacuum representation does not contain any
charged state!
} 
The vector state is guaranteed to be pure
only relative to the algebra of all bounded operators on $\Hz$.
But this algebra is far bigger than the algebra $\Az$, and contains
lots of operators that have no interpretation as q-observables.
This is the essential difference to the case of type I algebras,
realizable in a Fock space, which have many pure states. These algebras
are the local algebras of free quantum field theories, and only
encode systems of noninteracting particles.

Thus what breaks down in quantum field theory is the simple equation
''q-observable = self-adjoint Hermitian linear operator''.
Once this equation is broken, the question whether a
state is pure becomes dependent on the precise specification of which
operators are q-observables. In gauge theories the situation is further
complicated by the fact that the local algebras have a nontrivial
center consisting of charges that in each irreducible representation
are represented trivially. Thus a single irreducible representation
on a single Hilbert space (corresponding to a single superselection
sector) is no longer sufficient to characterize the complete algebra
of local q-observables.

To give up the assumption that every bounded self-adjoint operator is
a q-observable has serious consequences for the interpretation of
quantum physics.
Indeed, a test for a pure state is in terms of q-observables an
observation of the orthogonal projector to the subspace spanned by the
state. If this is not a q-observable then it is in principle impossible
to make this test. Thus testing for being in a pure state is impossible,
since these are no longer physical propositions.

So one cannot decide whether a system is in a pure state. To assume
this is thus a metaphysical act, and one can dispense with it without
any loss of information. But then the conventional form of Born's rule
hangs in the air, and all traditional interpretations break down
completely, since they start with Born's rule for pure states and
derive everything else from it.

\section{Requirements for good foundations}\label{s.req}

\nopagebreak
\hfill\parbox[t]{14.6cm}{\footnotesize

{\em The ordinary language, (spiced with technical jargon for the sake 
of conciseness) is thus inseparably united, in a good theory, with 
whatever mathematical apparatus is necessary to deal with the 
quantitative aspects. It is only too true that, isolated from their 
physical context, the mathematical equations are meaningless: but if 
the theory is any good, the physical meaning which can be attached to 
them is unique.}

\hfill Leon Rosenfeld, 1957 \cite[p.41]{Ros}
}

\bigskip

\nopagebreak
\hfill\parbox[t]{14.6cm}{\footnotesize

{\em I feel induced to contradict emphatically an opinion that Professor
L. Rosenfeld has recently uttered in a meeting at Bristol, to the 
effect that a mathematically fully developed, good and self-consistent 
physical theory carries its interpretation in itself, there can be no 
question of changing the latter, of shuffling about the concepts and 
formulae.}

\hfill Erwin Schr\"odinger, 1958 \cite[p.170]{Schroedinger1958}
}

\bigskip

\nopagebreak
\hfill\parbox[t]{14.6cm}{\footnotesize

{\em A great physical theory is not mature until it has been put in a
precise mathematical form, and it is often only in such a mature form 
that it admits clear answers to conceptual problems.}

\hfill Arthur Wightman, 1976 \cite[p.158]{Wig.Hilb}
}

\bigskip

\subsection{Foundations independent of measurement}

\nopagebreak
\hfill\parbox[t]{14.6cm}{\footnotesize

{\em to me it must seem a mistake to permit theoretical description to 
be directly dependent upon acts of empirical assertions}

\hfill Albert Einstein, 1949 \cite{Einstein1949}
}

\bigskip

Measurement should not figure in the foundations of physics;
the case for this was vividly made by \sca{Bell} \cite{Bell.against}.
The analysis in Section \ref {s.critique} indeed shows that actual
measurement practice is in conflict with the traditional foundations,
due to a far too idealized view of measurement.

Thus we are lead to inquire how foundations independent of measurement
could look like. This can be studied by looking at the modern account 
of the oldest of the physical sciences, Euclidean geometry.

For Euclidean geometry, considered as a branch of physics, there is a
complete consensus about how theory and reality correspond (on
laboratory scales).

One first defines the corresponding calculus and names the quantities
that can be calculated from quantum mechanical models (or models of
the theory considered) with the appropriate names from experimental
geometry. Thus, initially, a circle was a material object with a round
shape, and the mathematical circle was an abstraction of these.

The Pythagoreans (and later Descartes and Hilbert with even more
precision) then developed a theory that gives a precise formal meaning
to all the geometrical concepts. This is pure mathematics, today encoded
in textbook linear algebra and analytic geometry.
The theory and the nomenclature were {\em developed} with the goal of
enabling this identification in a way consistent with tradition.

Starting with Plato, the theory took precedence, defining the perfect
concept. What was found in reality was viewed as an approximate,
imperfect realization of the theoretical concept.

This was done by declaring anything in real life resembling
an ideal point, line, plane, circle, etc., to be a point, line, plane,
circle, etc., if and only if it can be assigned in an approximate way
(determined by the heuristics of traditional measurement protocols,
whatever they are) the properties that the ideal point, line, plane,
circle, etc., has, consistent to the assumed accuracy with the
deductions from the theory. If the match is not good enough, we can
explore whether an improvement can be obtained by modifying measurement
protocols (devising more accurate instruments or more elaborate
error-reducing calculation schemes, etc.) or by modifying the theory
(to a non-Euclidean geometry, say, which uses the same concepts but
assumes slightly different properties relating them).

No significant philosophical problems are left; lucent, intuitive, and
logically impeccable foundations for Euclidean geometry have been
established in this way. This indicates the maturity of Euclidean
geometry as a scientific discipline.

Thus once the theory is mature, the identification with real life is
done in terms of the formal, purely mathematical theory developed,
giving an interpretation to the theory. In this way, physics inherits
the clarity of mathematics, the art and science of precise concepts
and relations.

In particular, with the Lagrangian and Hamiltonian formulations,
classical mechanics has also reached the status of maturity, and hence
is perceived by most physicists as clear and philosophically
unproblematic.

\bigskip

To define quantum physics (or any other physical theory) properly,
including a logically impeccable interpretation, one should therefore
proceed as in Euclidean geometry and classical mechanics.

One first needs to define the corresponding calculus; this has already
been done in Section \ref{ss.postulates}. Then one has to name
the quantities that can be calculated from quantum mechanical
models (or models of the theory considered) with the appropriate names
the experimental physicists use for organizing their data.

One can then develop a theory that gives a precise formal meaning to the
concepts physicists talk about. This is pure mathematics -- the
shut-up-and-calculate part of quantum physics.
Finally, an \bfi{interpretation} to the theory, i.e., the identification
with real life, is given {\em in terms of the formal theory developed}  
by means of 

(CC) {Callen's criterion:} 
Operationally, a system is in a given state if its properties are 
consistently described by the theory for this state.

This generalizes the way H.B. Callen justified phenomenological
equilibrium thermodynamics in his famous textbook (\sca{Callen}
\cite{Cal}), where he writes on p.15:
{\it
''Operationally, a system is in an equilibrium state if its properties
are consistently described by thermodynamic theory.''} 
At first sight, this sounds like a circular definition (and indeed
Callen classifies it as such). But a closer look shows there is no
circularity since the formal meaning of 'consistently described
by thermodynamic theory' is already known. The operational definition
simply moves this formal meaning from the domain of theory to the
domain of reality by defining when a real system deserves the
designation 'is in an equilibrium state'. In particular, this
definition allows one to determine experimentally whether or not a
system is in equilibrium.

For quantum physics, Callen's criterion implies that we declare 
anything in real life resembling
an ideal photon, electron, atom, molecule, crystal, ideal gas, etc.,
to be a photon, electron, atom, molecule, crystal, ideal gas, etc., if
and only if it can be assigned in an approximate way (determined by
the heuristics of traditional measurement protocols, whatever that is)
the properties that the ideal photon, electron, atom, molecule,
crystal, ideal gas, etc., has, consistent to the assumed accuracy with
the deductions from the theory.

This identification process is fairly independent of the way
measurements are done, as long as they are capable to produce the
required accuracy for the matching, hence carries no serious
philosophical difficulties.

Of course, any successful theory must be crafted in such a way that it
actually applies to reality -- otherwise the observed properties cannot
match the theoretical description.
On the other hand, as the quote by Callen emphasizes,
{\bf we already need the theory to define precisely what it is that we
observe}.

As a result, theoretical concepts and experimental techniques
complement each other in a way that, if a theory is reaching maturity,
it has developed its concepts to the point where they are a good match
to reality. We then say that:

{\bf (R)}
Something in real life 'is' an instance of the theoretical concept
if it matches the theoretical description sufficiently well.

It is not difficult to check that this holds not only in physics but
everywhere where we have clear concepts about some aspect of reality.

If the match between theory and observation is not good enough, we can
explore whether an improvement can be obtained by modifying measurement
protocols (devising more accurate instruments or more elaborate
error-reducing calculation schemes, etc.) or by modifying the theory
(to a hyper quantum physics, say, which uses the same concepts but
assumes slightly different properties relating them).

Having established informally that the theory is an appropriate
model for the physical aspects of reality, one can study the
measurement problem rigorously on this basis:
One declares that a real \bfi{detector} (in the sense of a complete
experimental arrangement including the numerical postprocessing of raw
results that gives the final result) performs a real \bfi{measurement}
of an ideal quantity if and only if the following holds:
Modeling the real detector as a macroscopic quantum system (with the
properties assigned to it by statistical mechanics/thermodynamics)
predicts raw measurements such that, in the model, the numerical
postprocessing of raw results that gives the final result is in
sufficient agreement with the value of the ideal quantity in the model.

Then measurement analysis is a scientific activity like any other
rather than a philosophical prerequisite for setting up a consistently
interpreted quantum physics. Indeed, this is the way high precision
experiments are designed and analyzed in practice.

\subsection{What is a measurement?}\label{s.whatM}

Good foundations including a good measurement
theory should be able to properly justify this informal consensus by
defining additional formal concepts about what constitutes measurement.
To be satisfying, these must behave within the theory just as their
informal relatives with the same name behave in reality.
Then instrument builders may use the theory to inform themselves of what
can possibly work, and instrument calibration assumes the laws of
physics to hold.

Thus measurement must be grounded in theory, not -- as in the 
traditional foundations -- the other way round! In complete 
foundations, there would be formal objects in the
mathematical theory corresponding to all informal objects discussed
by physicists, including those used when designing and performing 
measurements. Only then talking about the formal objects
and talking about the real objects is essentially isomorphic.
We are currently far from such complete foundations.

To understand the precise meaning of the notion of measurement we look
at measurement in the context of classical physics and chemistry.
There are two basic kinds of measurements, destructive measurements and
nondestructive measurements.

\bfi{Nondestructive measurements} either leave the state of the object
measured unchanged (such as in the measurement of the length of a
macroscopic object) or modify it
temporarily during the measurement (e.g., temporarily deforming it to
measure the stiffness) in such a way that the object returns to its
original state after the measurement is completed.
\bfi{Destructive measurements} permanently change the state of the
object measured, usually by destroying all or part of it during the
measurement process. Examples are the determination of the age of an
archeological artifact by dendrochronology, or many traditional methods
of finding the chemical composition of a material.

In both cases, the measurement gives some \bfi{posterior} information
about the \bfi{prior} state, i.e., the state of the object before the
start of the measurement process. In case of destructive measurements,
it also gives some information about the products of the destruction,
from which properties of the prior states are deduced by reasoning.

A characteristic context of destructive measurements is the presence of
a large, sufficiently homogeneous object. Tiny parts of it are subjected
to destructive measurements to discover their relevant properties.
The homogeneity of the object then implies that the properties deduced
from the destructive measurements are also properties of the remainder
of the object. Thus destructive measurements of a tiny fraction of a
homogeneous object give information about the whole object, including
its unmeasured part. By our definition, we obtain in this way a
nearly nondestructive measurement of the whole object.

Alternatively, a large number of essentially identical objects are
present, a few of which are subjected to a destructive measurement.
The results of the measurement are then taken as being representative
of the properties of the unmeasured objects.
In case the measurements on the objects measured do not agree, one can
still make statistical statements about the unmeasured objects,
approximately valid within the realm of validity of the law of large
numbers. However, this no longer gives valid information about a single
unmeasured object, but only information about the whole population of
unmeasured objects. Thus one may regard the measurement on multiple
trial objects as a measurement of the state of the whole population.

Based on this analysis we conclude:

{\bf (M)}
A property $P$ of a system $S$ (encoded in its state) has been measured
by another system, the \bfi{detector} $D$, if at the time of completion
of the measurement and a short time thereafter (long enough that the
information can be read by an observer) the detector state carries
enough information about the state of the measured system $S$ at the
time when the measurement process begins to deduce with sufficient
reliability the validity of property $P$ at that time.

\subsection{Beables}\label{ss.beables}

\nopagebreak
\hfill\parbox[t]{14.6cm}{\footnotesize

{\em The scientist [...] appears as {\bf realist} insofar as he seeks 
to describe a world independent of the acts of perception; as 
{\bf idealist} insofar as he looks upon the concepts and theories as 
the free inventions of the human spirit (not logically derivable from 
what is empirically given); as {\bf positivist} insofar as he considers 
his concepts and theories justified only to the extent to which they 
furnish a logical representation of relations among sensory experiences.
He may even appear as {\bf Platonist} or {\bf Pythagorean} insofar as 
he considers the viewpoint of logical simplicity as an indispensable 
and effective tool of his research.} ~~~[original {\bf bold} preserved]

\hfill Albert Einstein, 1949 \cite{Einstein1949} 
}

\bigskip

One of the basic problems with the traditional interpretations of
quantum mechanics is the difficulty to specify precisely what counts
as real. The physics before 1926 was explicitly
about discovering and objectively describing the true, reliably
repeatable features of nature, seen as objectively real.

After the establishment of modern quantum physics, the goal of physics
can (according to the traditional interpretations of quantum mechanics)
only be much more modest, to systematically describe what physicists
measure.
Nonetheless, physics continues to make objective claims about reality
that existed long before a physicist performed the first measurement,
such as the early history of the universe, the composition of distant
stars and galaxies of which we can measure not more than tiny specks of
light, the age of ancient artifacts dated by the radio carbon method.
Physics also makes definite statements about the distant future of our
solar systems -- independent of anyone being then around to measure
it.

Thus there is a fundamental discrepancy between what one part of
physics claims and what the traditional interpretations of quantum
mechanics allows one to claim. This discrepancy was discussed in a
paper by \sca{Bell} \cite{Bell}, where he writes that quantum mechanics
{\it ''is fundamentally about the results of 'measurements',
and therefore presupposes in addition to the 'system' (or object) a
'measurer' (or subject). [...] the theory is only approximately
unambiguous, only approximately self-consistent. [...] it is interesting
to speculate on the possibility that a future theory will not be
intrinsically ambiguous and approximate. Such a theory could not be
fundamentally about 'measurements', for that would again imply
incompleteness of the system and unanalyzed interventions from outside.
Rather it should again become possible to say of a system not that such
and such may be observed to be so but that such and such be so.
The theory would not be about 'observables' but about 'beables'.
These beables [...] should, on the macroscopic level, yield an image of
the everyday classical world, [...] the familiar language of everyday
affairs, including laboratory procedures, in which objective properties
-- beables -- are assigned to objects.''}

To find beables we note that the problems created by quantum mechanics
are absent in classical mechanics. Thus it seems that classical objects
\bfi{exist} in the sense that they have objective properties that
qualify as beables. The classical regime
is usually identified with macroscopic physics, where length and time
scales are long enough that the classical approximation of quantum
mechanics is accurate enough to be useful. This suggests that we look
at the visible parts of quantum experiments.

Most experiments done to probe the foundations of quantum physics are
done using optical devices. In quantum optics experiments, both sources
and beams are extended macroscopic objects describable by quantum field
theory and statistical mechanics. For  example, a laser beam is simply
a coherent state of the quantized electromagnetic field, concentrated
in a neighborhood of a line segment in space.

The sources have properties independent of measurement, and the beams
have properties independent of measurement. These are objects described
by quantum field theory. For example, the output of a laser (before or
after parametric down conversion or any other optical processing) is a
laser beam, or an arrangement of highly correlated beams. These are in
a well-defined state that can be probed by experiment. If this is done,
they are always found to have the properties ascribed to them by the
preparation procedure. One just needs sufficient time to collect the
information needed for a quantum state tomography. The complete state
is measurable in this way, reproducibly, to any given accuracy.

Neither the state of the laser nor of the beam is changed by one or more
measurements at the end of the beam. Moreover, these states can be
found to any desired accuracy by making sufficiently long and varied
measurements of the beam; how this is done is discussed in quantum
optics under the name of quantum tomography.

Thus these properties exist independent of any measurement -- just as
the moon exists even when nobody is looking at it. They can be
found through diligent measurement, just as properties of distant stars
and galaxies. They behave in every qualitative respect just like
classical properties of classical objects.

On the other hand, measuring individual particles is an erratic affair,
and unless experiments are specially tuned (''non-demolition
measurements'') they change the state of the individual particles
in an unpredictable way, so that -- like in a classical destructive
measurement -- their precise state before measurement can never be
ascertained. Only probabilities for their collective behavior
can be given by averaging over many observations of different
realizations.

For example, the analysis of experimental particle collisions is based
on measuring the momentum and charge of many individual collision
products. But {\em individual} collision events (the momentum and
charge of the individual collision products) are not predicted by the
theory -- only the possibilities and their collective statistics, their
distribution in a {\em collection} of equally prepared events.
Indeed, probabilities mean nothing for a single collision.
What does it mean that the particular collision event recorded at a
particular time in a particular place is obtained with probability 0.07?
Nothing at all; the single simply happened and has no associated
probability. A statement about probabilities is always a statement
about a process that can be repeated many times under essentially
identical conditions.

\bigskip

Thus we have found a class of beables: the densities, intensities, and
correlation functions used to describe optical fields. And we have found
a class of non-beables: the individual particles.
The beables are computable from quantum statistical mechanics and
quantum field theory as expectations -- not as eigenvalues of
q-observables!
They are associated with quantum statistical mechanics and
quantum field theory on the level of fields -- not on the
level of individual particles.

Therefore, sources and beams are much more real than particles.
The former, not the latter, must be the real players in solid
foundations.

The goal of fundamental physics to understand things at the smallest
scales possible resulted in quantum field theory, where fields, not
particles, play the fundamental role. Particles appear only as
asymptotic excitations of the fields. Quantum field theory (or perhaps
an underlying theory such as string theory of which quantum field theory
is an approximation) is supposed to determine the behavior at all
scales, and hence lead to insights into the world at large.

Indeed, as we have seen, the basis of our perception of an objective
reality is statistical mechanics and field theory -- not few-particle
quantum mechanics! The inappropriate focus on the particle aspect of
quantum mechanics created the appearance of mystery; common sense is
restored by focussing instead on the field aspect.

The deeper reason for this is that from a fundamental point of view,
the particle concept is a derived, approximate concept derived from the
more basic concept of an interacting relativistic quantum field theory.
The quantum particle concept makes mathematical sense only under very
special circumstances, namely in those where a system actually behaves
like particles do -- when they can be considered as being essentially
free, as before and after a scattering event in a dilute gas or a
particle collider ring. The particle concept makes intuitive sense
only under the same circumstances. (This is discussed in more detail in 
Part III \cite{Neu.IIIfound} of this series of papers.)

We conclude from our discussion that

{\bf (F)}
Fields are real and have associated beables, given by expectation
values.

Conclusion (F) is one of the basic assumptions of the thermal
interpretation of quantum physics to be discussed in Part II
\cite{Neu.IIfound} of this series of papers.
It is routinely used in equilibrium and nonequilibrium
statistical thermodynamics (see, e.g., \sca{Calzetta \& Hu}
\cite{CalH.book}).

However, quite early in the history of modern quantum physics,
\sca{Ehrenfest} \cite{Ehr} found a clean and exact relation between the
dynamical laws of classical mechanics and quantum mechanics.
The \bfi{Ehrenfest equation}\footnote{
It is a pity that Ehrenfest did not develop this equation to the point
where it would have amounted to an interpretation of quantum physics.
This could have avoided a lot of the subsequent confusion.
} 
states that
\lbeq{e.Ehrenfest}
\frac{d}{dt}\<A\>_t = \<H\lp A\>_t,
\eeq
where $\<\cdot\>_t$ is the expectation in the state at time $t$ and
\lbeq{e.lp}
H\lp A:=\frac{i}{\hbar}[H,A].
\eeq
The implication for the interpretation of quantum physics (to be 
discussed in Part II \cite{Neu.IIfound}) seem to have gone unnoticed 
in the literature, probably because of the very strong tradition that 
placed an unreasonable notion of quantum measurement at the very basis 
of quantum physics.

\subsection{What is a particle?}\label{ss.particle}

The preceding featured beams of light, conventionally associated with
massless particles, the photons, and showed that the beams themselves
are far more real than the photons that they are suposed to contain.

An independent indication of the unreal status of photons is provided by
an analysis of the photoelecric effect, that faint coherent laser
light falling on a photosensitive plate causes randomly placed
detection events following a Poisson distribution.  Conventionally,
this effect is ascribed to the particle nature of light, and each
detection event is taken as a proof that a photon arrived. Upon closer
analysis, however, the detection events are found to be artifacts
caused by the quantum nature of the photosensitive plate. This must
be the case because the analysis can be done in a model of the process
in which no photons exist. Such an analysis is done, e.g., in
Sections 9.1-9.5 of \sca{Mandel \& Wolf} \cite{ManW}, a standard
reference for quantum optics. It is a proof that detection events
happen in the detector without photons being present. Hence one cannot
tell from a detection event whether the cause was a photon or a
classical field. But if the detectors cannot even distinguish an
external classical field from an impinging photon in the theoretical
analysis, based on which analysis should an experimentor decide?

In the model, the electron field of the
detector responds to a classical external electromagnetic radiation
field by emitting electrons according to Poisson-law probabilities.
Thus the quantum detector produces discrete Poisson-distributed clicks,
although the source is completely continuous. The state space of this
quantum system consists of multi-electron states only. Thus the
multi-electron system (followed by a macroscopic decoherence process
that leads to the multiple localization of the emitted electron
field) is responsible for the creation of the discrete detection
pattern.\footnote{
This was already clearly expressed in 1924 by \sca{Jeans} \cite{Jea},
who writes on p. 80: ''The fundamental law of quantum-dynamics, that 
radiant energy is emitted and absorbed only in complete quanta, is no 
longer interpreted as meaning that the ether can carry radiant energy 
only in complete quanta, but that matter can deliver or absorb radiant 
energy only by complete quanta.''
} 

Quantum electrodynamics is of course needed to explain special
quantum effects of light revealed in modern experiments, but not for
the photoelectric effect. Indeed, finer analysis reveals that beams in
nonclassical states may give a counting statistics significantly
different from that of the classical analysis. But this only shows that
the beam description needs quantum field theory (where people
conventionally use the language of photons), not that there must be
actual particles called photons.

It may seem, however, that the reality of individual {\em massive}
particles is established beyond doubt through the observation of
particle tracks in bubble chambers and other path-tracking devices.
But are the observed ''tracks'' guaranteed to be traces of particles?

The paper by \sca{Schirber} \cite{Schir} discusses essentially the same
phenomenon in a fully classical context, where a bullet is fired into a
sheet of glass and produces a large number of radial cracks in random
directions. This is shown in the first figure there, whose caption says,
{\it ''The number of cracks produced by a projectile hitting a glass
sheet provides information on the impactor's speed and the properties
of the sheet.''} In the main text, we read
{\it ''A projectile traveling at 22.2 meters per second generates four
cracks in a 1-millimeter-thick sheet of Plexiglas. [...]
A 56.7 meter-per-second projectile generates eight radial cracks
in the same thickness Plexiglass sheet as above.''}
(See also \sca{Falcao \& Parisio} \cite{FalP} and \sca{Vandenberghe \&
Villermaux} \cite{VanV}.)

We see that the discrete, random detection events (the cracks) are a
manifestation of broken symmetry when something impacts a material that
-- unlike water -- cannot respond in a radially symmetric way.
Randomness is inevitable in the breaking of a radial symmetry into
discrete events. The projectile creates an outgoing spherical stress 
wave in the plexiglas and produces straight cracks. In fact, once i
nitiated, the growth of a crack in a solid is not very different from 
the growth of a track in a bubble chamber, except that the energies 
and time scales are quite different. Only the initiation is random.

Would observed tracks in a high energy collision energy experiment
prove without doubt the existence of particles, one would have to
conclude that the projectile contains ''crack particles'' whose number 
is a function of the energy of the projectile -- 
just as the number of photons in a laser beam is a function of
its energy, and the number of events produced by a laser beam hitting a
photodetector provides information on the impactor's brightness.
Only the details are different.

Therefore the number of discrete detection events cannot be regarded
as obvious evidence for the existence of the same number of associated
invisible objects. They are at best evidence of the impact of something.

How do we know whether the tracks in a bubble chamber do not have a
similar origin? In both cases (bullet tracks and tracks in a bubble
chamber), something impinging on the detector produces a collection of
lines or curves. While the details are different, the mechanism is the
same. In each case one has a macroscopic and very complicated process
that breaks the symmetry and produces tracklike events. Thus there is
no a priori reason why in one case but not the other the lines should be
interpreted as evidence of particles.

It is strange that tradition says that in the classical experiment
with the bullet we see broken symmetry due to microscopic uncertainty,
whereas in a bubble chamber we see irreducible quantum randomness.

Tracks in a bubble chamber
are also a manifestation of broken symmetry when a radially symmetric
alpha-particle field produced by a radioactive nucleus impacts a
bubble chamber. A famous paper by \sca{Mott} \cite{Mott} (see also
\sca{Figari \& Teta} \cite{FigT,DelFT})
explains in detail how in a bubble chamber complete particle tracks
appear in random directions because of the discrete quantum nature of
the bubble chamber -- nowhere is made use of the particle nature of the
impacting radial wave! Thus while the details are quantum mechanical,
the underlying principle is classical!

What we see in a bubble chamber are droplets condensing due to 
ionization caused by a local piece of a spherical wave emanating from a 
radioactive nucleus. Mott analyzes the impact of the spherical wave and 
proceeds without reference to anything outside the quantum formalism.
He shows (p.80) that, in the absence of a deflecting magnetic field, 
the atoms cannot both be ionized unless they lie in a nearly straight 
line with the radioactive nucleus. There is no direct reference to the 
$\alpha$ particle causing the ionizations. 

Mott's analysis suggests that after the collision the scattered part 
forms a spherical wave (and not particles flying in different 
directions) until the wave reaches the detector. The spherical wave is 
nowhere replaced by flying particles. This makes his analysis very 
close to a field theoretical treatment. Particles appear to be 
ghostlike and only macroscopic (hence field-like) things are observed.

Mott needs Born's rule only for interpreting the final outcome in terms 
of probabilities and finds it consistent with a distribution of 
straight path only. This fully explains the tracks, without making any 
claims about position measurements or particle pointer states or
collapse assumptions. 

\bigskip

Everything said in this and the prededing subsection supports the view
that, except during the detection event, particles are unreal in the
sense of having no associated beables, and that they have a shade of
reality only as identical realizations of a population.

This conclusion that particles are unreal and have meaning only during
measurement or figuratively as part of a population is also reflected
in the statistical interpretation of quantum mechanics championed by
\sca{Ballentine} \cite[Chapter 9]{Ball.book}, who denies that a single
system has a state, and instead assigns the state to a population of
similarly prepared systems:  He writes on p.240: 
{\it ''regard the state operator $\rho$ as the fundamental description
of the state generated by the thermal emission process, which yields a
population of systems each of which is a single electron''}.
Thus he gives reality (beable properties) to the preparation procedure,
-- to the beam --, but not to the electrons (which do not even have
a state).

Slightly more indirectly, this is also reflected in the Copenhagen
interpretation of quantum mechanics, some versions of which 
say\footnote{
This goes back to \sca{Heisenberg}\cite{Hei1927}, who states on p.176:
{\it ''However, a single photon of such light is enough to eject the 
electron completely from its 'path' (so that only a single point of 
such a path can be defined). Therefore here the word 'path' has no 
definable meaning.''}
( {\it ''Von solchem Licht aber gen\"ugt ein einziges Lichtquant, um 
das Elektron v\"ollig aus seiner 'Bahn' zu werfen (weshalb von einer 
solchen Bahn immer nur ein einziger Raumpunkt definiert werden kann), 
das Wort 'Bahn' hat hier also keinen vern\"unftigen Sinn.''})
And on p.185 he states that {\it the 'path' comes into being only when 
we observe it}'' ({\it ''Die 'Bahn' entsteht erst dadurch, da{\ss} wir 
sie beobachten''}).
}
that an unmeasured system has no position, hence (in terms of beables) 
is unreal.

We conclude that the lack of reality of quantum particles is well
supported by the literature, in different ways. In contrast to (F) from 
Subsection \ref{ss.beables}, and in view of the Copenhagen 
interpretation, we may rephrase our findings as follows:

{\bf (P)}
Particles do not exist except when measured. They are detection
events created by the detector and mediated by fields.

What exists are the beams, and the discrete effects they produce when
subjected to measurement. This is an intuitive picture completely
orthogonal to the traditional interpretations of quantum mechanics.

It is a historical accident that one continues to use the name particle
in the many microscopic situations where it is grossly inappropriate if
one thinks of it with the classical meaning of a tiny bullet moving
through space. Restrict the use of the particle concept to where it is
appropriate, or do not think of particles as ''objects'' -- in both
cases all mystery is gone, and the foundations become fully rational.

\section{Outlook: New foundations}\label{s.outlook}

In the present paper, the universally accepted formal core of quantum
physics was cleanly separated from the controversial interpretation
issues. Moreover, it was shown that Born's rule, usually considered
a findamental, exact property of nature, has -- like most other law of
 nature -- its limitations, and cannot be regarded as a fundamental law.

Based on this insight, the other parts of this series of papers
present a new view of the foundations for quantum mechanics (QM) and
quantum field theory (QFT). This section puts the results of
Parts II--V \cite{Neu.IIfound,Neu.IIIfound,Neu.IVfound,Neu.Vfound}
into perspective, pointing to a coherent quantum physics where classical
and quantum thinking live peacefully side by side and jointly fertilize
the intuition.

\subsection{The thermal interpretation}\label{ss.therm}

The insight that Born's rule has its limitations and hence cannot be
the foundation of quantum physics opens the way for an alternative
interpretation -- the thermal interpretation of quantum physics,
defined and discussed in detail in Part II \cite{Neu.IIfound},
Part III \cite{Neu.IIIfound}, and Part IV \cite{Neu.IVfound}.
It gives new foundations that connect quantum physics (including
quantum mechanics, statistical mechanics, quantum field theory and
their applications) to experiment.

The new foundations improve the traditional foundations in several
respects:

\pt
The thermal interpretation is independent of the measurement problem.
The latter becomes a precise problem in statistical mechanics rather
than a fuzzy and problematic notion in the foundations.

\pt
The thermal interpretation better reflects the actual practice of
quantum physics, especially regarding its macroscopic implications.

\pt
The thermal interpretation explains the peculiar features of the
Copenhagen interpretation (lacking realism between measurements) and the
statistical interpretation (lacking realism for the single case) in the
microscopic world where the latter apply.

\pt
The thermal interpretation explains the emergence of probabilities from
the linear, deterministic dynamics of quantum states or quantum
observables.

\pt
The thermal interpretation gives a fair account of the interpretational
differences between quantum mechanics and quantum field theory.

\subsection{Questioning the traditional foundations in other respects}
\label{ss.questioning}

Traditionally, those learning quantum theory are expected to abandon
classical thinking and to learn thinking in a quantum mechanical
framework completely different from that of classical mechanics.
Though students widely differ in the order in which this happens, 
sooner or later, most of them are introduced to the items mentioned in 
the following caricature. 

\begin{itemize}
\item
Typically, they are first told that the Bohr--Sommerfeld theory of
quantization gave (for that time) an exact explanation of the spectral
lines for the hydrogen atom, firmly establishing that Nature is
quantized.
\item
Then they are told that Bohr's view is obsolete, and that it was just
a happy (or even misleading) coincidence that the old quantum theory
worked for hydrogen.
\item
Therefore, they are next acquainted with wave functions on
configuration space, their inner product, and the resulting Hilbert
spaces of square integrable wave functions. But almost immediately,
unnormalizable wave functions are used that do not belong to the
Hilbert space.
\item
They are then told with little intuitive guidance (except for a vague
postulated correspondence principle that cannot be made to work in many
cases) that in quantum mechanics, observables are replaced by Hermitian
operators on this Hilbert space.
\item
Later they may (or may not) learn that many of these operators are not 
even defined on the Hilbert space but only on a subspace.
\item
In particular, they are told that particles have no definite position
or momentum, but that these miraculously get random values when
measured, due to a postulated collapse of the wave function that
prepares the system in a new pure position or momentum state (which
does not exist since it is unnormalized).
\item
Now they must swallow a mysterious law defining the distribution of
these random values, called Born's rule. It is justified by the remark
that it is proved by (the Stern--Gerlach) experiment. But Born's rule
is claimed to hold for all conceivable quantum measurements, although
this experiment neither demonstrates the measurement of position nor
of momentum or other important quantities.
\item
They must then learn that between measurements, position and momentum
and hence well-defined paths do not exist. This leaves unexplained how
the Stern--Gerlach screen can possibly find out that a particle arrived
to be measured. The Stern--Gerlach device and all of quantum mechanics
begins to look like magic.
\item
Then they are taught the connection to classical physics by establishing
the Ehrenfest theorem for expectation values that obey approximately
classical laws. Although well-defined paths do not exist, the system
miraculously has at all times a well-defined mean path, even when
not measured.
\item
As a result, classical and quantum physics appear like totally
separated realms with totally different concepts and tools, connected
only by a rough-and-ready notion of correspondence that is ambiguous
and never made precise but works in a few key cases (and always with
liberally enough usage).
\item
After considerable time, when they have some experience with spectral
calculations, they learn how to use group theory (or, for those with
only little algebra background, spherical harmonics -- rotation group
representation tools in disguise) to determine the spectrum for
hydrogen. Miraculously, the results are identical with those obtained
by Bohr.
\item
At a far later stage, they meet (if at all) coherent states for
describing laser light, or as a tool for a semiclassical understanding
of the harmonic oscillator. Miraculously, a coherent state happens to
perform under the quantum dynamics exact classical oscillations.
\item
Only few students will also meet coherent states for the hydrogen atom,
the Berry phase, Maslov indices, and the accompanying theory of
geometric quantization, which gives the (slightly corrected)
Bohr--Sommerfeld rules for the spectrum a very respectable place in the
quantum theory of exactly solvable systems, even today relevant for
semiclassical approximations.
\item
Even fewer students realize that this implies that, after all,
classical mechanics and quantum mechanics are not that far apart.
A development of quantum mechanics emphasizing the closeness of
classical mechanics and quantum mechanics can be found
in my online book, \sca{Neumaier \& Westra} \cite{NeuW}.
\end{itemize}

Why does the conventional curriculum lead to such a strange state of
affairs? Perhaps this is the case because tradition builds the quantum
edifice on a time-honored foundation which accounts for essentially
all experimental facts but takes a ''shut-up-and-calculate'' attitude
with respect to the interpretation of the foundations. The traditional
presentation of quantum physics is clearly adequate for prediction but
seems not to be suitable for an adequate understanding.

\subsection{Coherent quantum physics}\label{ss.cohQM}

\nopagebreak
\hfill\parbox[t]{8cm}{\footnotesize

{\em Are coherent states the natural language of quantum theory?}

\hfill John Klauder, 1986 \cite{Kla.nat}
}

\bigskip

The conundrums of Subsection \ref{ss.questioning} are settled in
Part V \cite{Neu.Vfound} through the development of the
concept of a coherent quantum physics that removes the radical split
between classical mechanics and quantum mechanics.

That this might be feasible is already suggested by the history of
coherent states. (For the early history of coherent states see, e.g.,
\sca{Nieto} \cite{Nie}.)
Already in 1926, at the very beginning of modern quantum physics,
coherent states were used by \sca{Schr\"odinger} \cite{Schr} to
demonstrate the closeness of classical and quantum mechanical
descriptions of a physical system. Schr\"odinger discussed the main
properties of the coherent states today known as Glauber coherent
states. He did not call them coherent states, a notion coined in 1963
by \sca{Glauber} \cite{Gla}. Today the term \bfi{coherent
state} denotes a variety of (in detail very different) collections of
states displaying simultaneously a classical and a quantum character.

Part V is also part of another series of papers
(see \sca{Neumaier} \cite{cohSpaces.html}) that develop a theory of
coherent spaces, in terms of which the notion of a Glauber coherent
state is further generalized. Together with the thermal interpretation
from Part II, it serves as the basis for new foundations for quantum
physics in which classical and quantum thinking live peacefully side
by side and jointly fertilize the intuition. The fundamental importance
of coherent states is emphasized
by defining a \bfi{coherent quantum physics}, based on the concept of
coherence in various forms, thereby rebuilding from scratch the
foundations of quantum physics. Summarizing the vision in the shortest
terms, we may say:

\pt
Coherent quantum physics is quantum physics in terms of a
coherent space consisting of a line bundle over a classical phase space
and an appropriate ''coherent product'' characterizing the physical
properties of a quantum system.

\pt
The kinematical structure of quantum physics and the meaning of the
q-observables are given by the symmetries of this coherent space.

\pt
The dynamics is given by von Neumann's equation for the density
operator.

\pt
The connection to experiment is given through the thermal
interpretation.

Coherent spaces reconcile the old (semiclassical, Bohr-style)
thinking with the requirements of the new (operator-base) quantum
physics. They become the foundation on which a better, coherent quantum
physics is built. Mathematically, these foundations are equivalent to
the traditional Hilbert space approach. But conceptually, these
foundations begin with what is common between the classical and the
quantum world.

The only miracles left in the new approach outlined in this paper are
of a linguistic nature, namely the coincidence that the same word
''coherence'' fits several different contexts that come together in the
new foundations:

\pt
spatio-temporal coherence, the meaning of the term in ''coherent
states'',

\pt
logical coherence, refering to mathematically sound foundations,

\pt
intuitive coherence, implying that concepts make holistic sense to the
intellect, and

\pt
coherence as harmony, the meaning of the term in ''coherent
configurations'' and ''coherent algebras'', concepts from the
combinatorics of symmetry.\footnote{
Coherent configurations and the associated coherent algebras were
introduced in pure mathematics around 1970 (by \sca{Higman}
\cite{Hig.cc1,Hig.ca}), completely independent of physical
considerations. Special classes of coherent configurations called
\bfi{association schemes} and \bfi{distance-regular graphs} are very
well-studied, and many interesting examples are known in detail.
\\
Coherent configurations are, in a sense made precise in the companion
paper \cite{Neu.cohFin}, closely related to a finite variant of
coherent states.
Just as semisimple Lie groups act as symmetry groups of associated
Riemannian symmetric spaces, and their representation theory leads to
Perelomov coherent states, so most finite simple groups act as symmetry
groups of associated distance regular graphs. The latter is recorded
in the book by \sca{Brouwer} et al. \cite{BroCN}), of which I am a
coauthor. The book appeared just about the time when, for completely
different reasons, I turned to seriously study quantum physics.
\\
Only much later, I realized the extent of the connection of that work to
coherent states. The two subjects (quantum physics and the
combinatorics of symmetry) met in the past only in one area, the study
of \bfi{symmetric, informationally complete, positive operator valued
measures} (\bfi{SIC-POVM}s); see, e.g., \cite{wiki.SIC-POVM}.
Today's main open problem in the study of SIC-POVMs is \bfi{Zauner's
conjecture}, which dates back to the 1999 Ph.D. thesis of \sca{Zauner}
\cite{Zau}, written under my supervision.
}

In the literature, one usually finds coherent states discussed just for
themselves, or in the context of the classical limit. However, they are
also a powerful instrument in other respects. The reason is
that they have both a good intuitive semiclassical interpretation and
give good access to the whole Hilbert space (and beyond).
Indeed, coherent quantum physics turns coherent states into the
fundamental tool for studying quantum physics.

\bigskip

\end{document}